\documentclass[%
prl,%
preprint,
aps,%
a4paper,%
superscriptaddress,%
floatfix%
]{revtex4-1}

\usepackage{graphicx}
\usepackage{bm}
\usepackage{amsmath}
\usepackage{amssymb}
\usepackage{float}
\usepackage[utf8]{inputenc}
\usepackage[T1]{fontenc}
\usepackage[USenglish]{babel}
\usepackage{microtype}
\usepackage{nicefrac}
\usepackage{xspace}
\usepackage[caption=false]{subfig}
\usepackage[color=black!10]{todonotes}

\usepackage[colorlinks,citecolor=blue,urlcolor=blue,linkcolor=blue,bookmarks=false,filecolor=blue,runcolor=blue]{hyperref}
\usepackage[all]{hypcap}
\newcommand{\figref}[1]{Fig.~\ref{#1}}
\newcommand{\intextcite}[1]{Ref.~\citenum{#1}}

\newcommand{\unit}[1]{\ensuremath{~\text{#1}}}

\renewcommand{\Im}{%
\mathrm{Im}
}
\newcommand{\GM}{\ensuremath{\Gamma\text{M}}\xspace}
\newcommand{\GK}{\ensuremath{\Gamma\text{K}}\xspace}
\newcommand{\MK}{\ensuremath{\text{M}\text{K}}\xspace}
\newcommand{\iX}{iX\xspace}
\newcommand{\rangstrom}[1]{
\ensuremath{#1\,\text{\AA}^{-1}}
}
\newcommand{\hbn}{h-BN\xspace}
\newcommand{\wien}{WIEN2k\xspace}
\begin{document}
\title{Direct Observation of the Lowest Indirect Exciton State in the Bulk of Hexagonal Boron Nitride}
\author{R. Schuster}
\email[Author to whom any correspondence should be addressed: ]{r.schuster@ifw-dresden.de}
\author{C. Habenicht}
\affiliation{IFW Dresden, Institute for Solid State Research, P.O. Box 270116, D-01171 Dresden, Germany}
\author{M. Ahmad}
\affiliation{IFW Dresden, Institute for Solid State Research, P.O. Box 270116, D-01171 Dresden, Germany}
\affiliation{Department of Physics, COMSATS Institute of Information Technology, Park Road, 45550 Islamabad, Pakistan}
\author{M. Knupfer}
\affiliation{IFW Dresden, Institute for Solid State Research, P.O. Box 270116, D-01171 Dresden, Germany}
\author{B. Büchner}
\affiliation{IFW Dresden, Institute for Solid State Research, P.O. Box 270116, D-01171 Dresden, Germany}
\affiliation{Department of Physics, Technische Universität Dresden, 01062 Dresden, Germany}
\date{\today}

\begin{abstract} 
We combine electron energy-loss spectroscopy and first-principles calculations based on density-functional theory (DFT) to identify the lowest indirect exciton state in the in-plane charge response of hexagonal boron nitride (\hbn) single crystals. This remarkably sharp mode  forms a narrow pocket with a dispersion bandwidth of $\sim 100\unit{meV}$ and, as we argue based on a comparison to our DFT calculations, is predominantly polarized along the \GK-direction of the hexagonal Brillouin zone. Our data support the recent report by \citet{Cassabois_NatPhoton_2016_v10_p262} who indirectly inferred the existence of this mode from the photoluminescence signal, thereby establishing \hbn as an indirect semiconductor.
\end{abstract}

\maketitle

Hexagonal Boron Nitride (\hbn) is the binary analog of graphite. Like its carbon counterpart, it consists of $sp^2$-hybridized hexagonal layers that are stacked along the crystallographic $c$-axis. This leads to many similarities between these two materials like the applicability as dry lubricants or the possibility to wrap the hexagonal sheets into nanotubes \cite{Rubio_Phys.Rev.B_1994_v49_p5081,Chopra_Science_1995_v269_p966}.

More recently, \hbn came into focus in the field of 2D semiconductors, either as a substrate to boost the performance of graphene devices \cite{Dean_NatNano_2010_v5_p722} or as an important ingredient in so called van der Waals heterostructures \cite{Geim_Nature_2013_v499_p419}.
These overall similarities between the carbon and BN-case notwithstanding, the electronegativity  difference between the boron- and nitrogen-atoms has profound implications which show up most prominently in the electronic structure, in particular the fundamental band gap. Though its apparent simplicity in terms of structure and electronic properties, there has been an ongoing argument about the nature and size of the band gap. For a long time, there has been a strong controversy about whether the fundamental gap  is direct or indirect and reported values on the size of the gap $E_G$ range from below $4\unit{eV}$ to more than $7\unit{eV}$, depending on the sample quality and the employed method  (see e.g. \cite{Solozhenko_J.Phys.Chem.Solids_2001_v62_p1331,Akamaru_J.Phys.Chem.Solids_2002_v63_p887} for extended compilations of available data on $E_G$). 

With time, in particular in the theoretical community, consensus emerged that the gap is indirect \cite{Catellani_Phys.Rev.B_1987_v36_p6105,Blase_Phys.Rev.B_1995_v51_p6868,Arnaud_Phys.Rev.Lett._2006_v96_p26402} but still, calculated values differed substantially between $3.9\unit{eV}$ and $5.95\unit{eV}$. These predictions were, however, strongly challenged by the remarkable observation of a strong photoluminescence (PL) signal in \hbn reported in 2004 \cite{Watanabe_NatMater_2004_v3_p404} which promised great potential for lasing in the deep ultraviolet. 

Recently,  \citet{Cassabois_NatPhoton_2016_v10_p262} took an important step forward in resolving this long-standing debate. They inferred – although only indirectly – the existence of the lowest indirect exciton (\iX in their nomenclature) at an energy of $5.955\,\unit{eV}$ from a careful analysis of detailed PL data thereby establishing \hbn as an \emph{in}direct semiconductor also from an experimental perspective. They also could show  that the presence of sharp excitonic absorption (and/or emission) in a material with an indirect band gap – which appears to contradict the traditional understanding  \cite{Elliott_Phys.Rev._1957_v108_p1384} – occurs as a consequence of substantial exciton-phonon coupling which was deduced from the agreement between the satellite structure observed in the PL data and the phonon spectrum derived from inelastic x-ray scattering (IXS) \cite{Serrano_Phys.Rev.Lett._2007_v98_p95503}.

Also the indirect charge response which, in principle should provide more direct access to indirect exciton modes, has been the subject of previous studies employing electron energy-loss spectroscopy (EELS) \cite{Tarrio_Phys.Rev.B_1989_v40_p7852} and IXS \cite{Fugallo_Phys.Rev.B_2015_v92_p165122,Galambosi_Phys.Rev.B_2011_v83_p81413}. These results established a rich structure of excitonic states throughout the Brillouin zone (BZ) but failed to detect the \iX state, presumably due to insufficient energy resolution. Motivated by this, in the following we investigate the momentum dependent optical properties of \hbn with an unprecedented energy resolution which allows us to detect the previously hidden \iX state and characterize its behavior.

EELS in transmission is a bulk-sensitive scattering technique whose cross-section is proportional to the  so called loss-function $L(\bm{q},\omega)=\Im(-1/\epsilon(\bm{q},\omega))$, with $\epsilon(\bm{q},\omega)$ the longitudinal momentum- and energy-resolved dielectric function \cite{Sturm_Z.Naturforsch._1993}. In the past it has been used to address numerous issues related to the momentum-dependent charge response in widespread condensed-matter systems (see \cite{Roth_J.ElectronSpectrosc.Relat.Phenom._2014_v195_p85,Fink_J.ElectronSpectrosc.Relat.Phenom._2001_v117-118_p287} for an overview). More specifically, it has proven useful for the investigation of excitonic states in (in)organic semiconductors \cite{Schuster_Phys.Rev.Lett._2007_v98_p37402,Roth_J.Chem.Phys._2012_v136_p204708,Roth_NewJ.Phys._2013_v15_p125024,Schuster_Phys.Rev.Lett._2015_v115_p26404}.  Furthermore, from a  Kramers-Kronig analysis (KKA) of the measured data, it is possible to retrieve the complete dielectric function $\epsilon(\bm{q},\omega)=\epsilon_1(\bm{q},\omega)+\mathrm{i}\epsilon_2(\bm{q},\omega)$ thereby providing the possibility to compare the obtained results to other variants of optical spectroscopy.

\begin{figure}[h]
  \centering
  \includegraphics[width=\columnwidth]{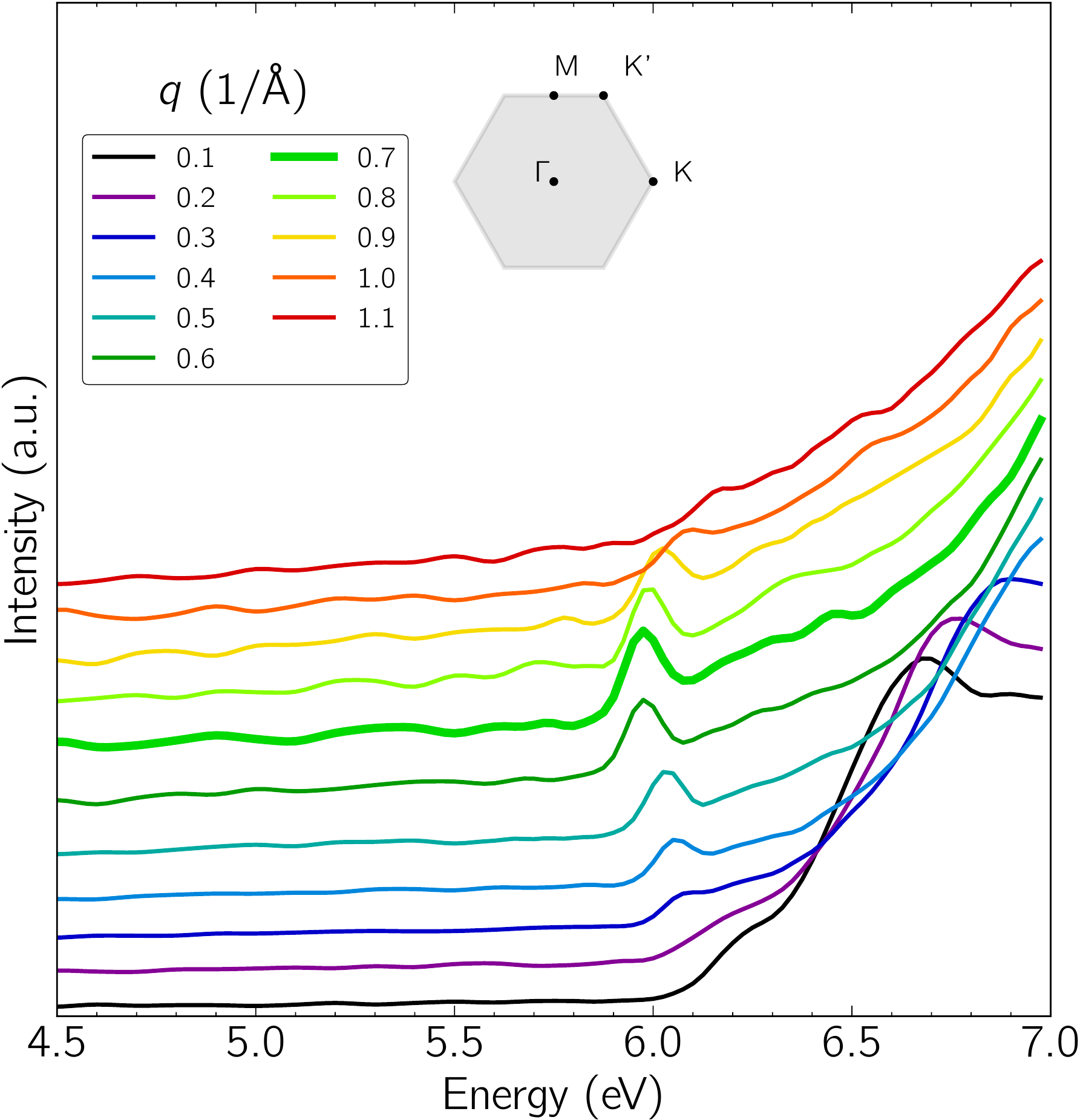}
  \caption{(Color Online) The EELS intensity in the region of the absorption onset measured parallel \GK for \hbn (for clarity the curves have been shifted vertically). Note the sharp mode that is visible in the momentum region between \rangstrom{0.4} and \rangstrom{1}. The thick line marks the momentum transfer where this mode is strongest and at its minimum energy position. The inset shows the hexagonal BZ illustrating that $q$ polarized along \GK is also sensitive to the symmetry-equivalent \MK-direction.}
\label{fig:fig1}
\end{figure}

For the present experiments, films were exfoliated from  single crystals purchased from “2d Semiconductors Inc.” \footnote{http://www.2dsemiconductors.com/}. The bulk-like films ($d\approx100-200\unit{nm}$) were put on standard electon-microscopy grids and then transferred to the spectrometer where they have been aligned \emph{in-situ}  with electron diffraction.  The measurements were carried out using a purpose-built transmission electron energy-loss spectrometer \cite{Fink_Adv.Electron.ElectronPhys._1989} with a primary electron energy of $172\unit{keV}$ and energy and momentum resolutions of $\Delta E = 80\unit{meV}$ and $\Delta q =\rangstrom{0.035}$, respectively and at temperatures of $T\approx20\unit{K}$ to minimize thermal broadening. All discussed features remain qualitatively unaltered at room temperature. We emphasize that \emph{a priori} the spectrometer setup does not allow to distinguish the contributions from symmetry-equivalent $q$-vectors within the scattering plane. For the specific case of \hbn, this means that the charge response along the \MK- and \GK-cut of the hexagonal Brillouin zone (BZ) are superimposed (see also the inset to \figref{fig:fig1}). In the following we use arguments based on a comparsion of our experimental data and DFT calculations to disentangle the contributions from these two symmetry-related directions.

\figref{fig:fig1} summarizes the EELS response for scattering parallel to the \GK direction of the hexagonal BZ. For small values of momentum transfer the spectrum in the vicinity of the fundamental gap edge is dominated by a peak at about $6.8\unit{eV}$ (which we ascribe based on our band structure calculations below to interband transitions into unoccupied states of predominant B $s$- and $p$-character) that sits on the low-energy tail of the $\pi$-plasmon (that peaks at $E\sim8.4\unit{eV}$) – the collective oscillation of the $\pi$-electron system \cite{Tarrio_Phys.Rev.B_1989_v40_p7852}. Upon increasing the momentum transfer, this mode disperses to higher energies and a sharp mode emerges that acquires its minimum energy ($E\sim5.98\unit{eV}$) and maximal spectral weight at about $q=Q_e=\rangstrom{0.7}$. Note that this feature is absent for $q\|\GM$. For still higher momenta – towards the BZ boundary – this mode disappears again by merging with the continuum of the higher-lying interband states. 

\begin{figure}[h]
  \centering
  \includegraphics[width=\columnwidth]{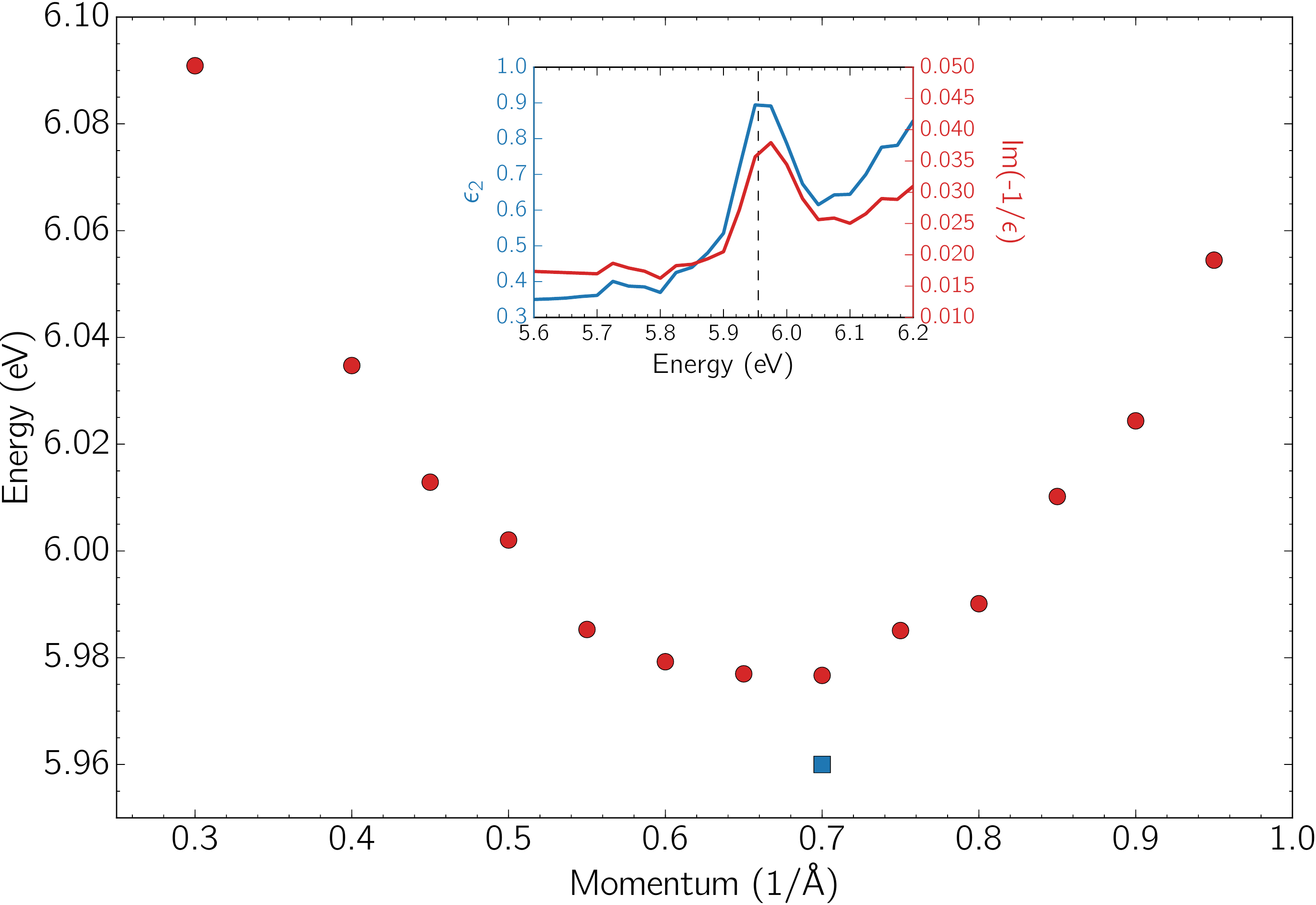}
  \caption{(Color Online) The dispersion of the low-energy mode. The circles are determined from the peak position in the measured EELS intensity shown in \figref{fig:fig1} whereas the square indicates the position of the corresponding peak in $\epsilon_2(\omega)$ at the minimum of the observed pocket at $q=Q_e$ as obtained from a KKA of the data (see inset and text for details). The dashed line in the inset corresponds to $E=5.955\unit{eV}$.}
\label{fig:fig2}
\end{figure}

To further elucidate the nature of the sharp mode we show its dispersion in \figref{fig:fig2}. Indeed, this feature forms a narrow pocket centered around $Q_e$ and exhibits a sizable bandwidth of about $100\unit{meV}$. This, together with the sharpness of this mode leads us to rule out absorption from valence band (VB) states into ionized impurity states below the band gap as a possible origin of this feature. 

It is well known that in general the loss function peaks at an energy blue-shifted with respect to $\epsilon_2(\omega)$ \cite{Kuzmany_SolidStateSpectroscopy_2009}. In order to facilitate a more straightforward comparison to conventional optical spectroscopy that usually probes the absorption coefficient $\alpha(\omega)\propto\epsilon_2(\omega)$, the data for $q=Q_e$ have been analyzed in terms of a KKA (see inset of \figref{fig:fig2}). Prior to this, the measured EELS intensity was corrected for contributions from the quasi-elastic line and multiple scattering and the KKA was normalized to yield results consistent with the $f$-sum-rule. From the derived shape of $\epsilon_2(\omega)$, we obtain $E=(5.96\pm0.06)\unit{eV}$ for the sharp mode which is in excellent agreement with the value of $E_{\text{\iX}}=5.955\unit{eV}$ reported in \cite{Cassabois_NatPhoton_2016_v10_p262} for the lowest indirect exciton state and we are therefore tempted to identify the mode we observe with the \iX-state.

Although this interpretation is appealing, it is not excluded that such a sharp absorption could be an effect caused by interband transitions that are not excitonic, i.e. bound in origin but are rather caused by excitations between the VB and conduction band (CB) states of the single-particle band structure. To investigate the possible role of these effects, in the following, we model the band structure of \hbn with density-functional theory (DFT).

\begin{figure}[h]
  \centering
  \includegraphics[width=\columnwidth]{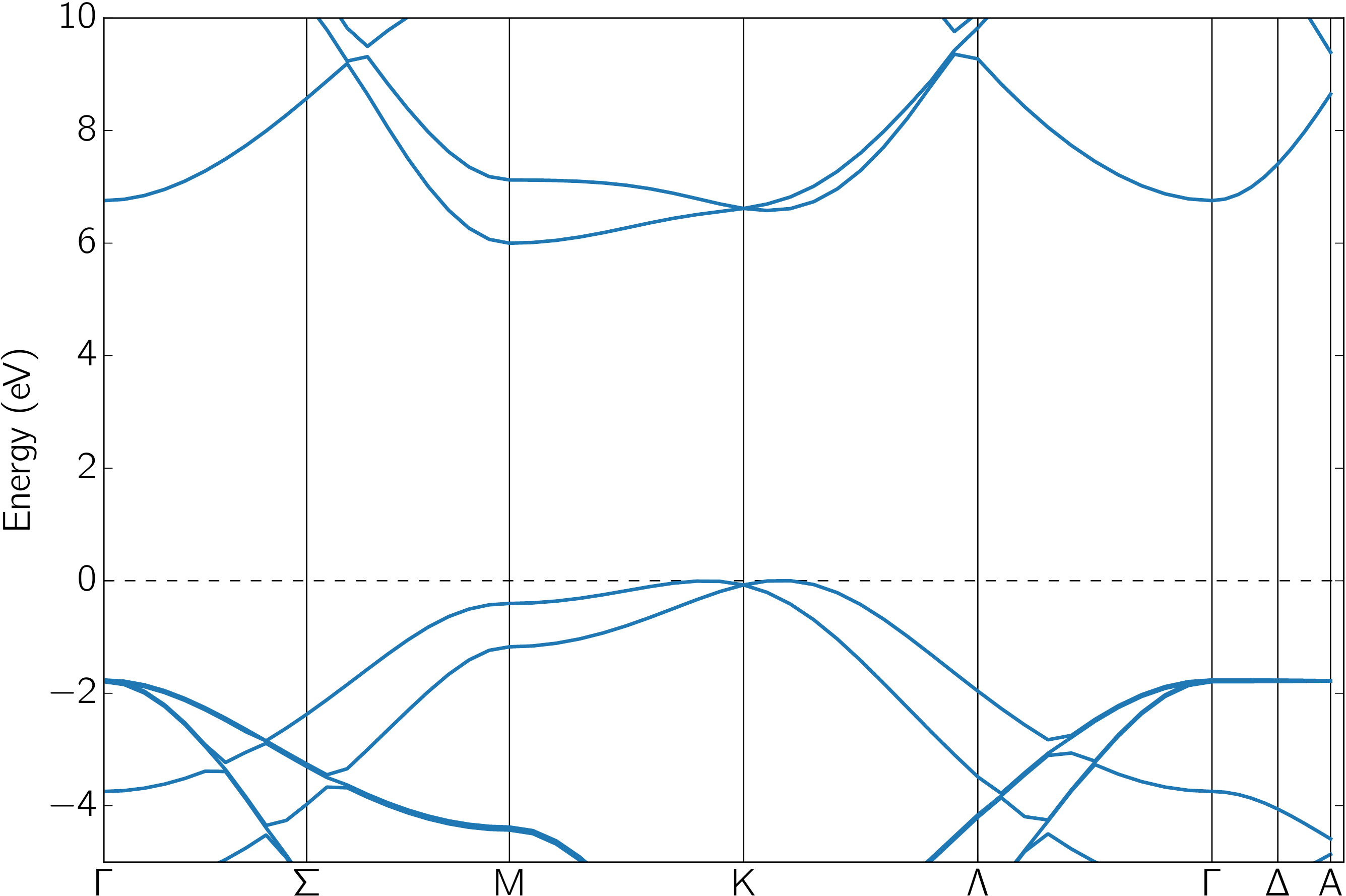}
  \caption{(Color Online) Band structure of \hbn as obtained from \wien employing the modified Becke-Johnson Potential. The zero of the energy axis is set to the VB maximum.}
\label{fig:band_structure}
\end{figure}

\hbn adopts a hexagonal lattice (space group 194: P6$_3$/mmc) with four atoms in the unit cell. For the first-principles calculations we took the experimental lattice constants of $a=b=2.51\unit{\AA}$ and $c=6.69\unit{\AA}$. All calculations have been performed with 
\wien \footnote{P. Blaha, K. Schwarz, G. K. H. Madsen, D. Kvasnicka and J. Luitz, WIEN2k, An Augmented Plane Wave + Local Orbitals Program for Calculating Crystal Properties (Karlheinz Schwarz, Techn. Universität Wien, Austria), 2001. ISBN 3-9501031-1-2}. The self-consistency cylce was converged on a $18\times18\times5$ grid of $k$-points and we adopted the generalized-gradient approximation (GGA) as suggested by \cite{Perdew_Phys.Rev.Lett._1996_v77_p3865} for the exchange correlation potential. To overcome the deficiency of GGA to correctly reproduce the band gap, we employed the modified Becke-Johnson (mBJ) semilocal exchange-correlation potential in its original version suggested in \intextcite{Tran_Phys.Rev.Lett._2009_v102_p226401} and implemented in \wien. The results of our band structure calculation are shown in \figref{fig:band_structure}. In good agreement with more elaborate $GW$-calculations \cite{Arnaud_Phys.Rev.Lett._2006_v96_p26402,Henck_Phys.Rev.B_2017_v95_p85410}, we obtain an indirect gap of $E_G=5.998\unit{eV}$ (for comparison, within GGA we obtain $E_G=4.236\unit{eV}$). Note that this value is also very close to the experimental estimate of $E_G=6.08\unit{eV}$ \cite{Cassabois_NatPhoton_2016_v10_p262}.

\begin{figure*}[h]
  \centering
\includegraphics%
[width=8.6cm]%
{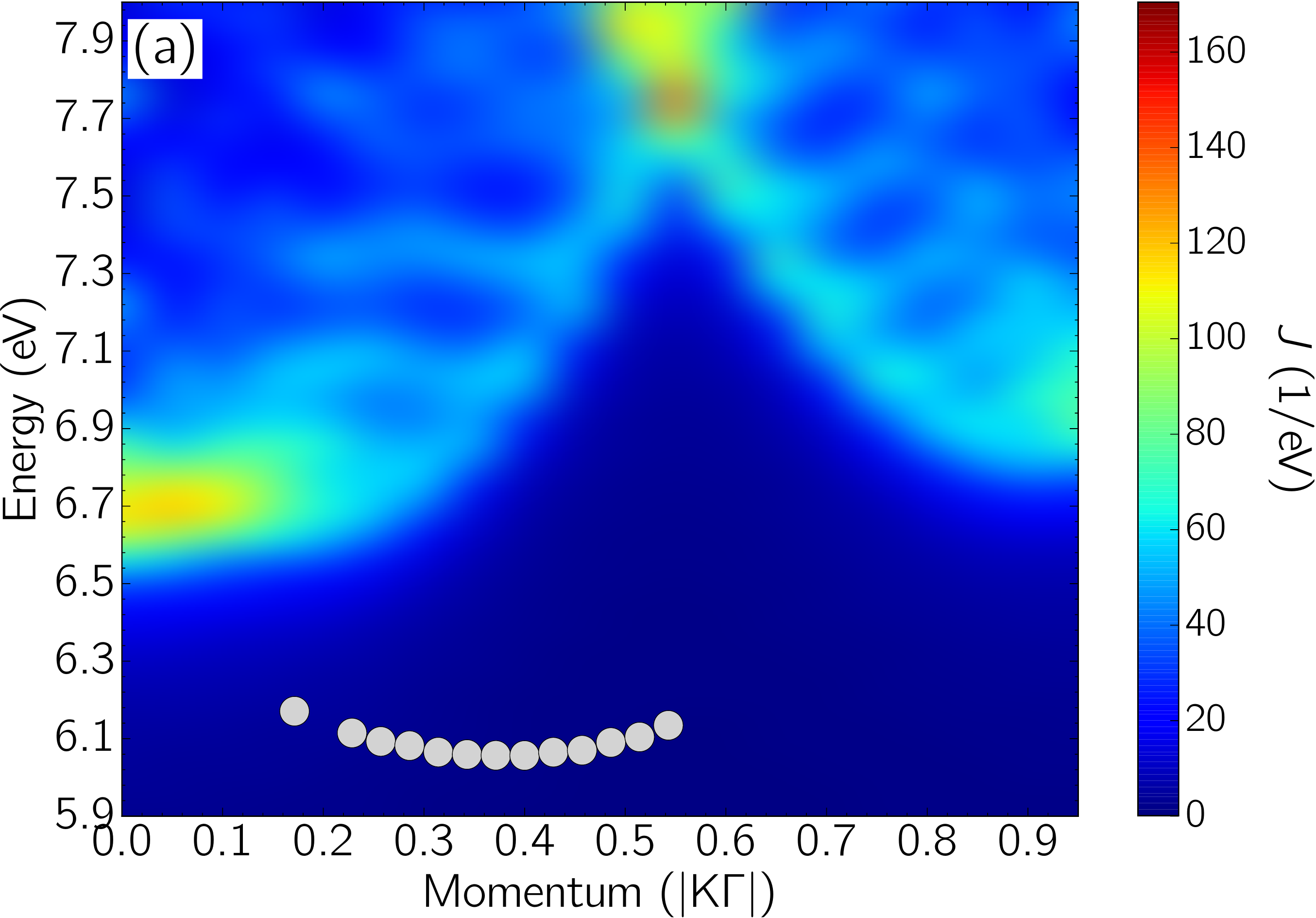}%
\hfill%
\includegraphics%
[width=8.6cm]%
{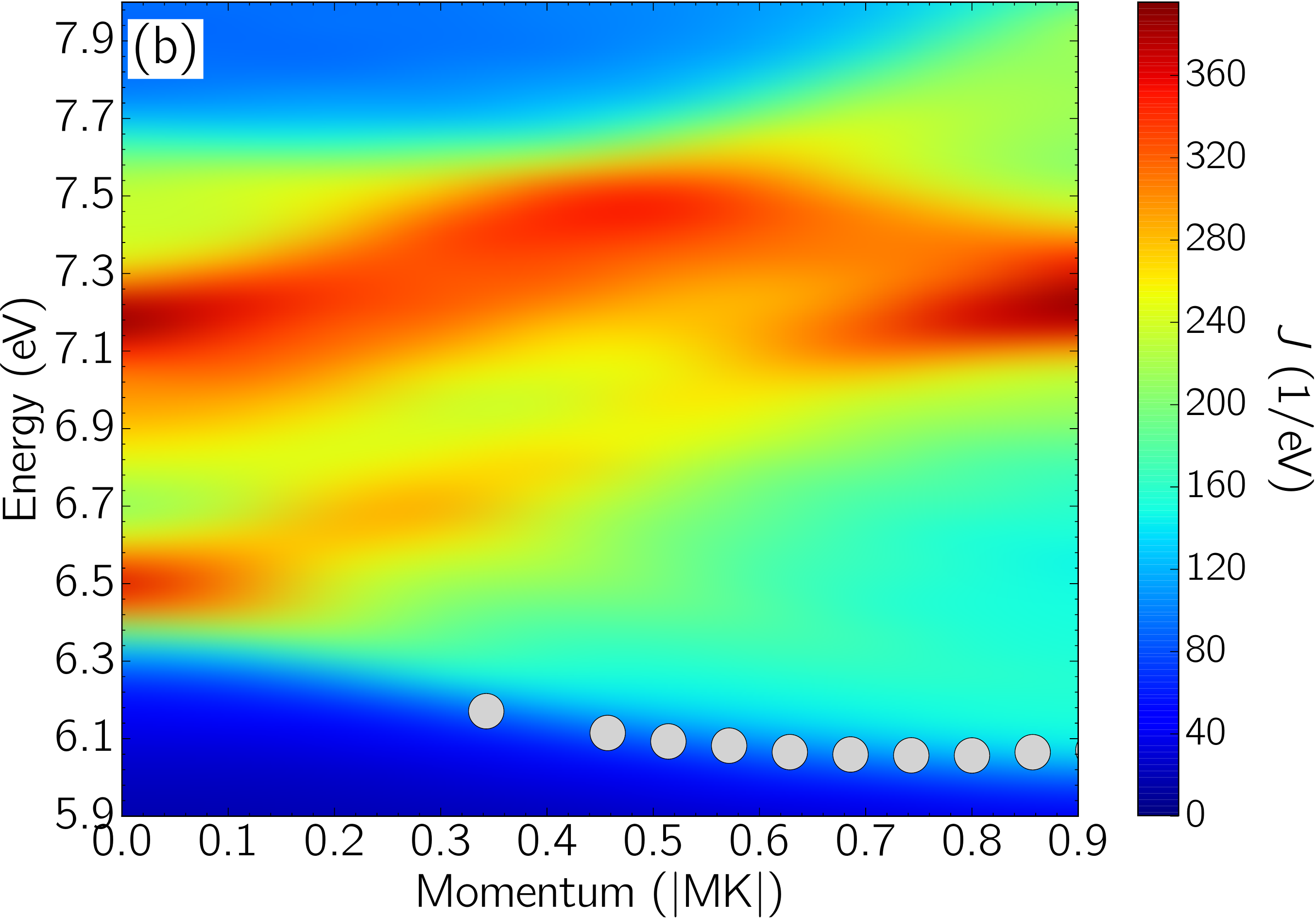}%
  \caption{(Color Online) Exciton \emph{vs.} the $q$-resolved JDOS from Eq.~\eqref{eq:1}. (a) The dispersion of the sharp mode shown in \figref{fig:fig2} against the background of possible single-particle transitions along the \GK-cut as calculated from the band structure shown in \figref{fig:band_structure}. Note that the exciton pocket is well defined with a binding energy of about $400\unit{meV}$ at $Q_e\sim 0.4\left|\text{K}\Gamma\right|$. (b) Same as (a) but now for the cut along \MK from \figref{fig:band_structure}. Note that in contrast to (a), the exciton lies inside the single-particle continuum. In both panels $J(q,\omega)$ from Eq.~\eqref{eq:1} has been smeared by $0.1\unit{eV}$.}
\label{fig:exciton_vs_jdos}
\end{figure*}

Besides a rigid energy shift of the eigenvalues, band curvatures are only modestly affected by this correction. This is also confirmed by the $GW$-predictions \cite{Blase_Phys.Rev.B_1995_v51_p6868,Arnaud_Phys.Rev.Lett._2006_v96_p26402,Henck_Phys.Rev.B_2017_v95_p85410} which are, however, computationally much more expensive and we therefore adhere to the mBJ-results in the remainder of the paper. 
To derive the momentum dependent optical properties from the DFT calculations beyond the dipole approximation, i.e. beyond the limit $q=0$, is a non-trivial task. Instead of calculating the full response function, we therefore define a $q$-resolved joint density of states (JDOS) $J(q,\omega)$ according to
\begin{equation}
  \label{eq:1}
  J(q,\omega)=\sum_{knn'}\,\delta\left(\omega-(\epsilon_{k+q,n'}-\epsilon_{k,n})\right)\,,
\end{equation}
which we take, in analogy to the $q=0$ case \cite{Grundmann_2006}, as an approximation for $\epsilon_2(q,\omega)$ or, equivalently, the absorption coefficient. In this equation the sum runs over (un)occupied (CB-) VB-states ($n'$) $n$ and the $\epsilon_k$ are the corresponding Kohn-Sham eigenvalues. This approach neglects the possible influence of matrix elements of the form $\left<\psi_{k+q,n'}\left|\text{e}^{\text{i}q\cdot r}\right|\psi_{k,n}\right>$ mediating the transitions from the initial into the final Bloch states. We justify this by the usually only small variations of the orbital projections of the involved states in the vicinity of the VB- and CB-edges. 
From Eq.~\eqref{eq:1} it is clear that for an interband transition to occur at a particular energy $\omega$ and momentum $q$, the energy separation between the VB-states and the $q$-shifted CB states must coincide with $\omega$. 
\figref{fig:exciton_vs_jdos} contains a comparison between the calculated results for $J$ and the experimentally observed dispersion already shown in \figref{fig:fig2} for the two relevant momentum cuts along \GK and \MK (see the discussion above). From this comparison, we observe that for momentum transfers along \GK, the sharp mode seen in the EELS response lies well below the single-particle continuum and therefore corresponds to an (indirect) exciton state. On the contrary, along the symmetry equivalent \MK cut, the mode lies inside the single-particle continuum  whose onset it traces very closely. This is at odds with the sharpness of this state, in particular for $q\sim Q_e$ where the presence of numerous decay channels from the continuum would suggest a much higher spectral width than what is observed in our experiments. We take this observation, together with the agreement in energy (see \figref{fig:fig2}), as strong evidence to identify the sharp state we observe in \figref{fig:fig1} and \figref{fig:fig2} with the lowest indirect exciton (\iX) state reported by \citet{Cassabois_NatPhoton_2016_v10_p262}. Based on a comparison of the experimentally observed spectral sharpness of this state with the expectation of substantial broadening due to relaxation into particle-hole transitions which we do expect along \MK, we argue that this exciton is predominantly polarized along the \GK direction of the BZ; which is in accord with the recent detection of the transition dipole lying in the BN-plane \cite{Vuong_2DMaterials_2017_v4_p11004}. Note that it is also only along this direction that the exciton acquires a substantial binding energy ($\sim400\unit{meV}$ at $q=Q_e$) which is in qualitative agreement with the predictions of possible exciton binding energies obtained from calculations based on the  Bethe-Salpeter equation (BSE) \cite{Arnaud_Phys.Rev.Lett._2006_v96_p26402}. Note, however, that these calculations were performed for a direct exciton state, i.e., for $q=0$, in contrast to our observations shown above. A sizable binding energy of the exciton is also needed in order to understand why we observe this mode even at room temperature. At present, it is not clear to us why 
the exciton decays outside the momentum window shown in \figref{fig:fig1} although the comparison to the JDOS would suggest a much larger phase space for the existence of the collective state (see \figref{fig:exciton_vs_jdos}(a)).

In summary, we investigated the indirect charge response of \hbn with electron energy-loss spectroscopy (EELS). We observe a sharp mode in a finite portion of the BZ which forms a narrow pocket centered around $Q_e\sim\rangstrom{0.7}$ with a minimum energy at about $E\sim5.955\unit{eV}$. Based on a comparison of our EELS data to predictions from first-principle calculations and recent observations reported in \cite{Cassabois_NatPhoton_2016_v10_p262} we identify this state as the lowest possible indirect exciton state in \hbn. Our data therefore confirms the observation of \hbn being an \emph{in}direct semiconductor. In addition, our observations call for further more elaborate \emph{ab-initio} studies of the exciton structure which, so far, did not predict the existence of this mode.

Our experimental work benefited from  technical support provided by R. Hübel, S. Leger and M. Naumann. M.A. acknowledges financial support of the German Academic Exchange Service (DAAD) under the Leibniz-DAAD post-doctoral fellowship program. C.H. and R.S. appreciate funding from the IFW excellence program. R.S. acknowledges computational hardware assistance by U. Nitzsche and enlightening discussions concerning the DFT calculations with R. Ray.
\bibliographystyle{apsrev4-1}

%

\end{document}